\documentclass[conference]{IEEEtran}

\IEEEoverridecommandlockouts
\usepackage{cite}
\usepackage{amsmath,amssymb,amsfonts}
\usepackage{graphicx}
\usepackage{textcomp}
\usepackage[dvipsnames]{xcolor}
\usepackage[hyphens]{url}  %

\def\BibTeX{{\rm B\kern-.05em{\sc i\kern-.025em b}\kern-.08em
    T\kern-.1667em\lower.7ex\hbox{E}\kern-.125emX}}

\usepackage{cite}
\usepackage{comment} %
\usepackage{amsthm} %
\usepackage{hyperref}
\usepackage{mathtools} %
\usepackage{amssymb} %
\usepackage{algorithm}
\usepackage{algpseudocode}
\usepackage{cleveref}
\usepackage{algorithmicx}
\usepackage{algpseudocode}

\usepackage{wrapfig}    %
\usepackage{caption}    %
\usepackage{subcaption} %
\usepackage{rotating}  %
\usepackage{booktabs} %
\usepackage{longtable} %
\usepackage{pdflscape} %
\usepackage{afterpage} %
\usepackage{multirow} %
\usepackage{subcaption} %
\usepackage{xcolor} %
\usepackage{xfrac} %
\usepackage[T1]{fontenc} %
\usepackage{dirtytalk} %
\usepackage{adjustbox} %
\usepackage{enumitem} %
\usepackage{array}
\usepackage{makecell}
\usepackage{float}  %
\usepackage{bbding} %
\usepackage{threeparttable} %

\def\Snospace~{\S{}}
\renewcommand*\sectionautorefname{\Snospace}
\renewcommand*\subsectionautorefname{\Snospace}

\hypersetup{
	pdftoolbar=true,        %
	pdfmenubar=true,        %
	pdffitwindow=false,     %
	pdfstartview={FitH},    %
	pdftitle={},    %
	pdfauthor={},     %
	pdfsubject={},   %
	pdfcreator={},   %
	pdfproducer={}, %
	pdfkeywords={}, %
	pdfnewwindow=true,      %
	colorlinks=true,       %
	linkcolor=Brown,          %
	citecolor=OliveGreen,        %
	filecolor=magenta,      %
	urlcolor=NavyBlue           %
}

\DeclareMathOperator*{\argmin}{arg\,min}
\newcolumntype{L}{>{\raggedright\arraybackslash}X}

\usepackage{orcidlink}

\newcommand{\todo}[1]{\textcolor{red}{#1}} %
\newcommand{\mypar}[1]{\smallskip\noindent\textbf{#1.}}

\newtheorem{definition}{Definition} %
\theoremstyle{definition}

\newcommand{\mycomment}[1]{} %

\newcommand{\Autoref}[1]{%
  \begingroup
    \renewcommand{\sectionautorefname}{Section}%
    \renewcommand{\subsectionautorefname}{Subsection}%
    \autoref{#1}%
  \endgroup
}

\begin{document}

\title{
Let’s Simply Count: Quantifying Distributional Similarity Between Activities in Event Data
}
\author{
	Henrik Kirchmann$^\dagger$\orcidlink{0009-0007-7521-2955}
	,
    Stephan A. Fahrenkrog-Petersen$^{\ddagger}$\orcidlink{0000-0002-1863-8390}.
        Xixi Lu$^\sharp$\orcidlink{0000-0002-9844-3330},
	Matthias Weidlich$^\dagger$\orcidlink{0000-0003-3325-7227
}\vspace{1em}\\
	\selectfont\rmfamily\itshape
	\small
	$^\dagger$Humboldt-Universit\"at zu Berlin, Germany\quad
    $^\ddagger$ University of Liechtenstein, Liechtenstein \quad
	$^\sharp$Utrecht University, The Netherlands\\
	\footnotesize \selectfont\ttfamily
    $^\dagger$firstname.lastname@hu-berlin.de\quad
	$^\sharp$x.lu@uu.nl\quad $\ddagger$ stephan.fahrenkrog@uni.li
}

\maketitle

\begin{abstract}

To obtain insights from event data, advanced process mining methods
assess the similarity of activities to incorporate their semantic relations
into the analysis.
Here, distributional similarity that captures similarity
from activity co-occurrences is commonly employed. However,
existing work for distributional similarity in process mining adopt
neural network-based approaches as developed for natural language processing,
e.g., word2vec and autoencoders. While these approaches have been shown to
be effective, their downsides are high computational costs and limited
interpretability of the learned representations.

In this work, we argue for simplicity in the modeling of distributional similarity of activities. We introduce count-based
embeddings that avoid a complex training process and offer a direct
interpretable representation. To underpin our call for simple embeddings, we
contribute a comprehensive benchmarking framework, which includes means to
assess the intrinsic quality of embeddings, their performance in downstream applications, and their computational efficiency. In experiments that
compare against the
state of the art, we demonstrate that count-based embeddings provide a
highly effective and efficient basis for distributional similarity between
activities in event data.

\end{abstract}


\section{Introduction}

Information systems across various industries generate large amounts of
event data while supporting business processes. Process mining aims at
transforming such event data into meaningful models,
insights, and actionable outcomes~\cite{van2016data}. Critical to this
transformation is an understanding of the event data in terms of the
activities for which the execution is signaled, and especially their
semantic relations.

Traditionally, activities in event logs have been treated as atomic,
symbolic tokens, equivalent to one‐hot representations that carry no notion
of similarity or distance. However, recent work has shown that approaches
which move beyond this assumption can substantially enhance a wide range of
process mining tasks, including trace clustering~\cite{bose2009context,
chiorrini2022embedding}, conformance
checking~\cite{peeperkorn2020conformance}, predictive
monitoring~\cite{camargo2019learning, gamallo2023learning}, anomaly
detection~\cite{nolle2022binet}, and process‐model repository
querying~\cite{wang2014querying}.

To capture the similarity of activities, many existing proposals adopt the
idea of distributional similarity~\cite{harris1954distributional}. That is,
similarity between activities is inferred directly from event data
by analyzing the distributional patterns in which activities occur.
Activities surrounded by similar preceding and succeeding steps are likely
to reflect related behavior, whereas differing contexts suggest distinct
process steps. %
The advantage of distributional similarity is that it can be derived
automatically and does not rely on scarce domain-specific background
knowledge. Rather, it models the relations between activities as they
directly emerge from the analyzed process.

Existing work for distributional similarity in process mining mainly
adopts neural network-based approaches, also referred to as prediction-based
approaches, as developed for natural languages,
e.g., word2vec and autoencoders.
Here, neural networks are trained to predict a words' context to
capture distributional information~\cite{levy2015improving}, thereby
avoiding an explicit encoding of co-occurrence information.
This way, dense, low-dimensional embeddings, i.e., vector representations,
are obtained even for natural languages with up to millions of distinct
words~\cite{jurafsky2025speech}. The
representations are not directly interpretable, though, and their construction
incurs high computational costs.

\begin{figure}[b!]
    \centering
    \includegraphics[width=\linewidth]{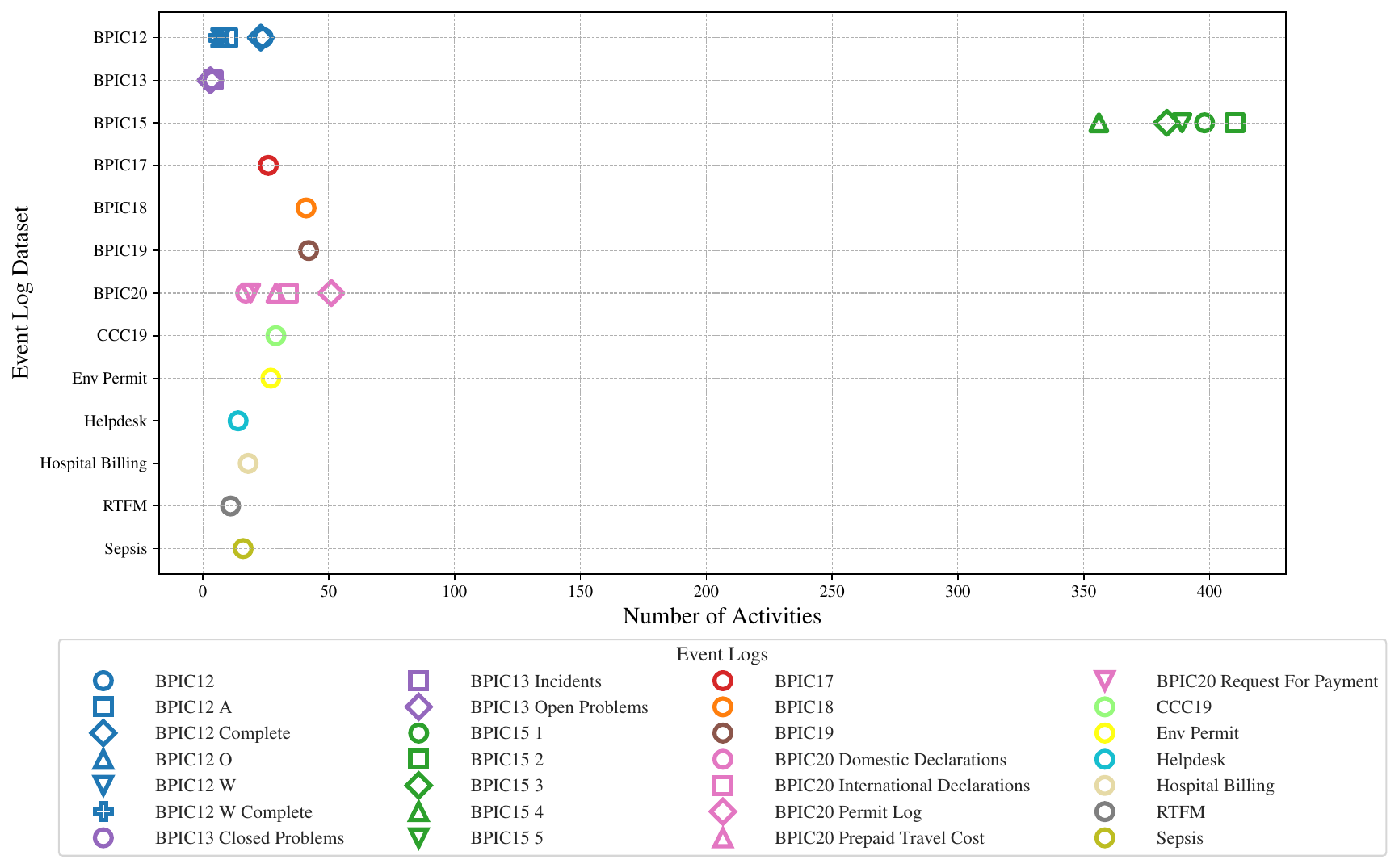}
    \caption{Number of unique activities in event logs.}
    \label{fig:number_of_activities}
\end{figure}

In this paper, we challenge the need for neural network-based approaches for
distributional similarity in process mining. In
fact, event logs involve much smaller vocabularies than natural languages.
Most processes contain tens to hundreds of unique
activities, as shown in \autoref{fig:number_of_activities} for
several real-world event logs. %
In such a setting, simple count‐based embeddings that encode co-occurrence
per individual activity become feasible. Unlike
neural network-based embeddings, they are intuitively
interpretable and can be derived efficiently. Despite these advantages, it
is not clear whether they are also effective for process mining use cases,
as no prior work systematically compared them for deriving activity
similarities.

Against this background, we provide a novel view on
distributional similarity of activities:
\begin{enumerate}[left=0.3em, nosep]
	\item We introduce count-based embeddings for activities in event
	data to capture distributional similarity. Those are based on a direct
	encoding of co-occurrences per activity.
	\item We present the first comprehensive benchmarking framework for
	evaluating methods for distributional similarity between activities.
	It includes:
	\begin{enumerate}[left=.3em]
		\item An intrinsic evaluation approach assessing similarity independently of downstream tasks.
		\item An evaluation of activity embeddings applied to the
		downstream task of next-activity prediction.
		\item A efficiency benchmark highlighting computational efficiency across methods.
	\end{enumerate}
\end{enumerate}
Using this benchmarking framework, we demonstrate that our count-based
approaches consistently outperform existing methods in terms of
effectiveness, while also excelling in computational efficiency.

Below, we first give formal preliminaries
(\autoref{sec:02_Background_and_Preliminaries})
and review related work (\autoref{sec:03_related_work}). We then introduce
our count-based embeddings (\autoref{sec:04_Count-Based Activity
Embeddings}). Next, we define and apply our benchmarking framework
(\autoref{sec:05_general_experimental_setup}-\autoref{sec:runtime}), before
presenting conclusions
(\autoref{sec:Conclusion}).

\section{Preliminaries} \label{sec:02_Background_and_Preliminaries}

\mypar{Sequence} For a given set \( A \), \( A^* \) is the set of all finite sequences over \( A \). A finite sequence over \( A \) of length \( n \) is a mapping \( \sigma \in \{1, \dots, n\} \to A \). Such a sequence is represented by a string, i.e., \( \sigma = \langle a_1, a_2, \dots, a_n \rangle \) where \( a_i = \sigma(i) \) for \( 1 \leq i \leq n \). 
We denote with \( \sigma[i,j] \) the subsequence \( \langle a_i, a_{i+1}, \dots, a_j \rangle \) of \( \sigma \) for any \( 1 \leq i \leq j \leq n \). The length of the sequence is denoted as \( |\sigma| \), i.e., \( |\sigma| = n \). The operation \( \sigma \oplus a' \) represents the sequence with element \( a' \) appended at the end, i.e., \( \sigma \oplus a' = \langle a_1, \dots, a_n, a' \rangle \). Similarly, \( \sigma_1 \oplus \sigma_2 \) appends sequence \( \sigma_2 \) to \( \sigma_1 \), resulting in a sequence of length \( |\sigma_1| + |\sigma_2| \). Moreover, for any element \( a \in A \), the notation \( a^n \) represents a sequence of length \( n \) consisting entirely of \( a \), i.e., \( a^n = \langle a, a, \dots, a \rangle \) with \( n \) occurrences of \( a \).  
For any sequence \( \sigma \) over \( A \), we define \( \partial_{\text{mset}}(\sigma) = [a_1, a_2, \dots, a_n] \). The operator \( \partial_{\text{mset}} \) converts a sequence into a multiset, preserving element multiplicities. For example, \( \partial_{\text{mset}}(\langle a, a, c, a, b \rangle) = [a^3, b, c] \).

\mypar{Event Data}
Let $\mathcal{A}$ be the universe of activities. A trace $t \in \mathcal{A}^*$ is a sequence of
activities. 
$\mathcal{T} = \mathcal{A}^*$ denotes the universe of traces.
Let $\mathcal{L} = \mathcal{B}(\mathcal{T})$ be the universe of event logs. An event log $L \in \mathcal{L}$ is a finite multiset of traces. Given an event log $L \in \mathcal{L}$, with $A_L \subseteq \mathcal{A}$ we denote the set of all activities in $L$. %

\section{Related Work} \label{sec:03_related_work}

\subsection{Modeling Similarity}  \label{subsec:Modeling Semantics}

Typically, one of the following three approaches is used to derive similarities between activities.

\mypar{Manual Definition Using Domain Knowledge} Process experts manually
specify distances between activities based on their knowledge of the process
\cite{montani2014retrieval}. This approach suffers from scalability issues
as it requires extensive manual effort, which becomes impractical for
processes that involve a high number of activities or are subject to change.
Also, it is limited in capturing complex or nuanced semantic relationships
\cite{jurafsky2025speech}.

\mypar{Inspecting Activity Labels} Natural language processing techniques
are used to extract similarities from textual descriptions of activity
labels \cite{rebmann2024evaluating, wang2014querying}. This approach faces
limitations when activity labels are encoded as non-natural language strings
(e.g., ``AWB45'') or when seemingly similar labels (e.g., ``Patient Hospital
Check-In'' and ``Patient Hospital Check-Out'') conceal fundamentally
different process steps.

\mypar{Distributional Similarity} This approach infers similarities directly from event data by analyzing the distributional patterns in which activities occur. Building on the distributional hypothesis \cite{harris1954distributional}, which holds that elements in similar contexts tend to share meaning \cite{firth1957synopsis}. We argue that the same principle applies to process activities. Activities surrounded by similar preceding and succeeding steps are likely to reflect related behavior, whereas differing contexts suggest distinct process steps. This approach captures activities as embeddings, latent high-dimensional vectors, where similar activities are mapped to nearby points. This enables capturing similarity without manual definitions or label interpretation.

\subsection{Distributional Similarity between Activities}
Existing work on distributional similarity between activity distances is limited to small‐scale, method‐specific evaluations with no unified benchmark. By contrast, trace‐level encodings and similarity measures have been systematically compared \cite{back2023comparing,tavares2023trace}. Below, we outline the existing methods we benchmark and their evaluation approaches.

\emph{Substitution Scores} \cite{bose2009context}:
This is the only existing method that is count-based; the authors derive activity similarities from log-ratios of the number of co-occurrences of activities. As we discuss in Section \ref{sec:04_Count-Based Activity Embeddings}, this approach is related to ours but differs in how activity similarities are derived.
The authors evaluate their approach on trace clustering with a small synthetic log, using activity similarities as substitution scores in an edit distance. Since edit distance also requires insertion costs, we do not replicate this benchmark.

\emph{act2vec} \cite{de2018act2vec}:
A neural network-based approach adapting Word2Vec \cite{mikolov2013efficient}, to derive activity embeddings,
but the activity embeddings are not evaluated.

\emph{Embedding Process Structure} \cite{chiorrini2022embedding}: A model-based method mapping each activity in a discovered Petri net to a six‐dimensional feature vector, incorporating features like path length, parallelism, or self-loopability.
Its proof-of-concept evaluation clusters activities across two versions of a loan‐application process, validated against the author's judgment; we omit this benchmark approach due to the single‐process scope and the lack of a ground truth.

\emph{Autoencoder} \cite{gamallo2023learning}:
A neural autoencoder learns activity embeddings by reconstructing context windows. These embeddings are evaluated as initialization for the embedding layer in next-activity prediction models, yielding better performance than random initialization. While comprehensive, the work does not compare against other embedding methods.

We conclude that count-based vector representations have not been used to
capture distributional similarity between activities, and we observe a lack
of systematic evaluations.

\section{Count-Based Activity Embeddings} \label{sec:04_Count-Based Activity Embeddings}

We present 12 different methods for generating activity embeddings by selecting one option from each of three design dimensions: two context interpretations (multiset or sequence), two matrix types (activity-activity co-occurrence or activity–context frequency), and three post-processing options (none, PMI or PPMI). Each method produces a matrix where each row represents the embedding of a single activity. Activity similarity is computed using the cosine distance between embeddings, with lower distances indicating higher similarity.

These design choices determine how the embeddings are generalized: each
choice affects how we model activities by their contexts or co-occurrences.
Finding the right balance is crucial for meaningful similarities.

\mypar{Context Interpretation}
As distributional similarity follows the idea of ``You shall know an
activity by the company it keeps,'' we first define this company, i.e., the
context. We extract it by sliding an \(n\)-length window over each trace
(see \autoref{alg:activity_context_extraction}). Each window is split into
left and right subwindows of size \(\lfloor\tfrac{n-1}{2}\rfloor\) and
\(\lceil\tfrac{n-1}{2}\rceil\). Concatenating these subwindows gives two
alternative context representations. \(C_{\mathrm{mset}}\) treats the
surrounding activities as a multiset, whereas \(C_{\mathrm{seq}}\) preserves
their order.
\mycomment{
\begin{algorithm}[H]
\caption{Activity Context Extraction}
\label{alg:activity_context_extraction}
\begin{algorithmic}[1]
    \State \textbf{Input:} Event log \( L \) and window size \( n \in \mathbb{N} \).
    \State Pad each trace \( t \in L \) to define the padded event log \( L_{\text{padded}} \):
    \[
    L_{\text{padded}} = [ \#^{\lfloor (n-1)/2 \rfloor} \oplus t \oplus \#^{\lceil (n-1)/2 \rceil} \mid t \in L ]
    \]
    where \( \# \notin A_L \) is a special padding symbol.

    \State Extract the set of all windows from \( L_{\text{padded}}\) of length $n$:
\[
\hspace{-0.247em}
W_n = \bigcup_{t \in L_{\text{padded}}} \big\{\! \langle a_i, a_{i+1}, \dots, a_{i+n-1} \rangle \mid 1 \leq i \leq |t| - n + 1 \!\big\}
\]
\State Define the set of all possible contexts:
\[
C_{lr} = \bigcup_{w \in W_n} \left\{ \!
\arraycolsep=2pt
\begin{array}{l|l}
(c_l, c_r) &
\begin{array}{l}
|c_l| = \left\lfloor \frac{n-1}{2} \right\rfloor,\quad
|c_r| = \left\lceil \frac{n-1}{2} \right\rceil, \\
c_l \oplus \langle a \rangle \oplus c_r = w,\quad
a \in A_L
\end{array}
\end{array}
\! \right\}
\]
\State Define the set of multiset and sequence context interpretations as: \[C_{mset} = \{ \partial_{\text{mset}}(c_l \oplus c_r)\ \mid  (c_l,c_r) \in C_{lr} \} \] \[C_{seq} = \{ c_l \oplus c_r\ \mid  (c_l,c_r) \in C_{lr} \} \]

\end{algorithmic}
\end{algorithm}
}

\mypar{Matrix Type}
Let \(C\in\{C_{\mathrm{mset}},C_{\mathrm{seq}}\}\) and \(\#(a,c)\) be the
count of activity \(a\in A_L\) in context \( c\in C\). We define:

\smallskip

\noindent\emph{Activity-Activity Co-occurrence Matrix} $AA: A_L\times A_L
\to \mathbb{N}$,
\[
AA(a,a')=\sum_{\substack{c\in C\\\#(a,c)>0,\;\#(a',c)>0}}
\#(a,c)+\#(a',c).
\]

\smallskip

\noindent\emph{Activity-Context Frequency Matrix} $AC: A_L\times C \to
\mathbb{N}$,
\[
AC(a,c)=\#(a,c).
\]

\scalebox{0.9}{
\begin{minipage}{\linewidth}
\begin{algorithm}[H]
\caption{Activity Context Extraction}
\label{alg:activity_context_extraction}
\begin{algorithmic}[1]
    \Require Event log \(L\), window size \(n\)
    \State Pad each trace $\sigma$, with padding symbol $\texttt{PAD}\notin A_L$:
      \[
        L' \gets \bigl[\texttt{PAD}^{\lfloor\frac{n-1}2\rfloor}\!\oplus \sigma\!\oplus\!\texttt{PAD}^{\lceil\frac{n-1}2\rceil}\mid \sigma\in L\bigr]
      \]
    \State Initialize \(C_{\mathrm{mset}}\leftarrow\emptyset,\;C_{\mathrm{seq}}\leftarrow\emptyset\)
    \ForAll{\(\sigma\in L'\)}
      \For{\(i=1\) to \(|\sigma|-n+1\)}
        \State \(w\gets \langle \sigma(i),\dots,\sigma(i+n-1)\rangle\)
        \State \( c_l\gets w[1:\lfloor\frac{n-1}2\rfloor],\;c_r\gets w[(\lfloor\frac{n-1}2\rfloor+2):n]\)
         \State $C_{\mathrm{mset}}\gets C_{\mathrm{mset}}\cup\{\partial_{\text{mset}}(c_l\!\oplus\!c_r)\}$
        \State $C_{\mathrm{seq}}\gets C_{\mathrm{seq}}\cup\{c_l\!\oplus\!c_r\}$
      \EndFor
    \EndFor
    \State \textbf{Output:} \(C_{\mathrm{mset}},\,C_{\mathrm{seq}}\)
\end{algorithmic}
\end{algorithm}
\vspace{.5em}
\end{minipage}
}

\mypar{Post-Processing: PMI \& PPMI}
Count‐based embeddings are skewed by very frequent but uninformative occurrences~\cite{levy2015improving}.
This phenomenon is well-known in natural language processing, where
high-frequency words, such as ``the'', ``a'', and ``have'', contribute
little to the true similarity to other words despite their frequent
(co-)occurrence.

\begin{figure}[b]
 	\centering
 	\includegraphics[width=\linewidth]{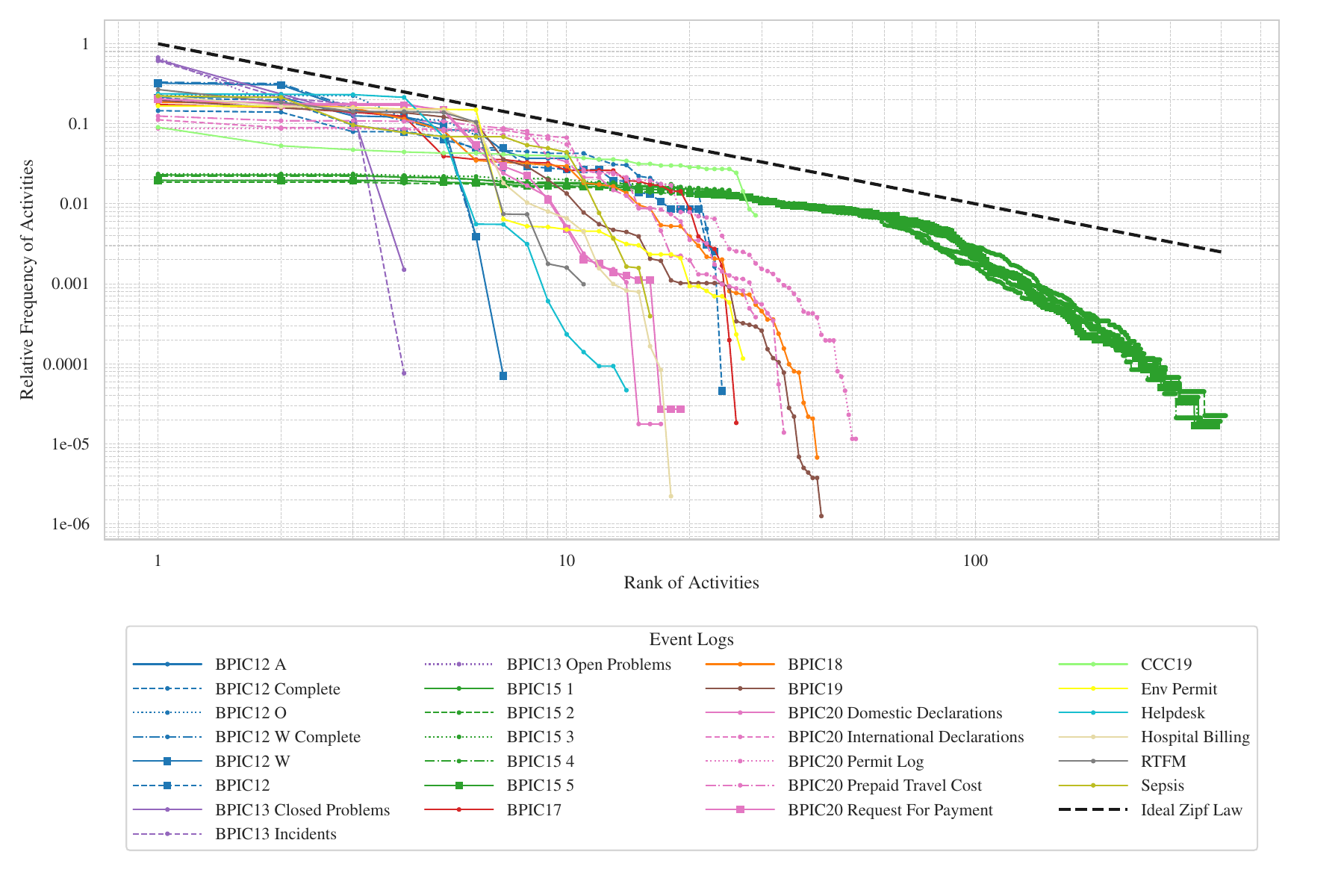}
 	\caption{Log–log plot of activity frequency vs.\ rank, compared to an ideal Zipf distribution.}
 	\label{fig:zipf}
 \end{figure}

Words in natural language follow Zipf’s law~\cite{piantadosi2014zipf}, which
states that the most common
word occurs approximately twice as often as the next common one, three times as often
as the third most common, and so on. \autoref{fig:zipf} shows that activity
frequencies in many logs match Zipf for top ranks but then fall more
steeply, making the frequency distribution of activities in event data also
highly skewed. A small number of activities occur very frequently, while
many others appear only rarely.

 To correct this, we use Pointwise Mutual Information (PMI)~\cite{church1990word} to quantify how much more often two events \(x\) and \(y\) co-occur than under independence.

Denote by \(\#(a)\) the number of occurrences of activity \(a\in A_L\), by \(\#(c)\) the number of occurrences of context \(c\in C\) in \(L\) and %
let \( N = \sum_{\sigma \in L} | \sigma | \). We define the probability of activity \(a\) and context \(c\) occurring in \(L\) as
\[
p(a)=\frac{\#(a)}{N},\qquad p(c)=\frac{\#(c)}{N}\,.
\]
Then
\[
AA^{\mathrm{PMI}}(a,a')=
\begin{cases}
\displaystyle
\log\frac{AA(a,a')/N}{p(a)\,p(a')}, & AA(a,a')>0,\\
0, & \text{otherwise},
\end{cases}
\]
\[
AC^{\mathrm{PMI}}(a,c)=
\begin{cases}
\displaystyle
\log\frac{AC(a,c)/N}{p(a)\,p(c)}, & AC(a,c)>0,\\
0, & \text{otherwise}.
\end{cases}
\]
PMI can yield negative values, which indicate less co-occurrence than expected by chance. Negative PMI values are typically unreliable, especially in smaller corpora \cite{jurafsky2025speech}, so we set them to zero, yielding Positive PMI (PPMI):
\[
AA^{\text{PPMI}}(a,a') \;=\;\max\bigl(0,\;AA^{\mathrm{PMI}}(a,a')\bigr),
\]
\[
AC^{\text{PPMI}}(a,c) \;=\;\max\bigl(0,\;AC^{\mathrm{PMI}}(a,c)\bigr).
\]

\mypar{Example} Let event log \( L = [\langle a,b,c,d,e\rangle^5,\, \langle a,d,d,b,e\rangle] \) and choose window size \( n = 3 \).
Then Algorithm~\ref{alg:activity_context_extraction} returns \( C_{\mathrm{mset}} = \{[\texttt{PAD}, b], [a,c], \dots\} \) and \( C_{\mathrm{seq}} = \{\langle \texttt{PAD}, b\rangle, \langle a, c\rangle, \dots\} \).
This yields the activity context-frequency matrix \( AC \)  for \( C_{\mathrm{mset}} \):

\[
\begin{array}{c|ccccccc}
& [\texttt{PAD}, b] & [a,c] & [b, d] & [c,e] & [d, \texttt{PAD}] & [a,d] & [d,e] \\\hline
a & 5 & 0 & 0 & 0 & 1 & 0 & 0 \\
b & 0 & 5 & 0 & 0 & 0 & 0 & 1 \\
c & 0 & 0 & 5 & 0 & 0 & 0 & 0 \\
d & 0 & 0 & 1 & 5 & 0 & 1 & 0 \\
e & 1 & 0 & 0 & 0 & 5 & 0 & 0
\end{array}
\]

Applying PMI yields for example \( AC^{\mathrm{PMI}}(c,[b,d]) \approx 1.43 \), \( AC^{\mathrm{PMI}}(d,[b,d]) \approx -0.34 \).

\mycomment{
\mypar{Example} Let event log $L$ be:
\[
L = [
  \langle a,b,c,d,e\rangle^5,\,
  \langle a,d,d,b,e\rangle\
]
\]
and choose window size \(n=3\). Algorithm~\ref{alg:activity_context_extraction} returns:
\begin{align*}
C_{\mathrm{mset}} &= \{
  [\texttt{PAD}, b], [a,c], [b, d], [c,e], [d, \texttt{PAD}], [a,d], [d,e] \}\\
C_{\mathrm{seq}} &= \{ \langle \texttt{PAD}, b \rangle, \langle a, c \rangle, \langle b,d\rangle, \langle c, e \rangle, \langle d,  \texttt{PAD}\rangle, \langle \texttt{PAD}, d \rangle,\\
&\quad \quad \langle a,d \rangle, \langle d, b\rangle, \langle b, \texttt{PAD}\rangle  \}
\end{align*}

The activity context-frequency $AC$ matrix for \( C_{\mathrm{mset}} \) is then defined as:

{\footnotesize
\[
\begin{array}{c|ccccccc}
& [\texttt{PAD}, b] & [a,c] & [b, d] & [c,e] & [d, \texttt{PAD}] & [a,d] & [d,e] \\\hline
a & 5 & 0 & 0 & 0 & 1 & 0 & 0 \\
b & 0 & 5 & 0 & 0 & 0 & 0 & 1 \\
c & 0 & 0 & 5 & 0 & 0 & 0 & 0 \\
d & 0 & 0 & 1 & 5 & 0 & 1 & 0 \\
e & 1 & 0 & 0 & 0 & 5 & 0 & 0
\end{array}
\]
}

Applying PMI to this matrix gives us for example:
\begin{align*}
AC^{\mathrm{PMI}}(c,[b,d]) &= \log\frac{5/30}{(6/30)\cdotp(6/30)} \approx 1.43 \\
AC^{\mathrm{PMI}}(d,[b,d]) &= \log\frac{1/30}{(7/30)\cdotp(6/30)} \approx -0.34  \\
\end{align*}
}

\mypar{Our Method vs. Substitution Scores} The substitution scores approach
\cite{bose2009context} is related to the PMI-weighted activity-activity
co-occurrence matrix with sequential context, differing mainly in the
denominator of the logarithmic function. While standard PMI uses \( p(a)
\cdot p(b) \) assuming independence, \cite{bose2009context} adjusts this to
\( p(a) \cdot p(b) \) when \( a = b \), and \( 2 \cdot p(a) \cdot p(b) \)
when \( a \neq b \). The rationale for this change is unclear, the
standard denominator reflects expected co-occurrence under independence
\cite{jurafsky2025speech}. Furthermore, we interpret the matrix rows as embeddings to derive similarity scores, whereas their method uses individual cells as pairwise scores, ignoring relationships with other activities.

\section{General Experimental Setup}
\label{sec:05_general_experimental_setup}

To evaluate the effectiveness and efficiency of methods for distributional
similarity between activities, we introduce a comprehensive benchmarking
framework. Here, we first outline the general experimental setup.

\mypar{Implementation \& Event Data}
Our code and event data are on
GitHub.\footnote{\url{{https://github.com/henrikkirchmann/semantic-aware-process-mining-distances}}}
Experiments ran on an Intel Xeon 6254, NVIDIA A100 (80 GB), and 756 GB RAM.
We use 28 real-world event logs that differ in various statistics like
number of unique activities (3–410), number of traces (20–251,734),
unique‐trace ratio (<0.01–1), or avg. trace length (2.87–69.7); any
exclusions are noted in the corresponding benchmark sections.

\mypar{Window Size}
For all methods, except the process‐model‐based approach \cite{chiorrini2022embedding} and the autoencoder‐based technique \cite{gamallo2023learning}, we evaluated window sizes of 3, 5, and 9. The process-model-based method has no window-size parameter, and, due to its long runtimes, we tested the autoencoder only at a window size of 3, following the original study’s finding that this setting yields the best predictive performance.

\mypar{Implementation Settings of Existing Approaches}
For all evaluated methods, we used their original implementation settings whenever available. The only exception is the autoencoder-based approach from \cite{gamallo2023learning}. While the original authors selected the best-performing model from a 5-fold cross-validation (using a 64/16/20 train/validation/test split), we opted for a single 80/20 train/validation split due to the high computational cost of repeated training. Additionally, since \cite{gamallo2023learning} does not specify a fixed embedding dimensionality, we set it to 128, aligning with the authors’ observation that larger vector sizes generally improve performance. For the act2vec approach~\cite{de2018act2vec}, we evaluate both the CBOW (multiset context) and skip-gram (sequence context) architectures.

\section{Intrinsic Evaluation} \label{sec:Intrinisc Evaluation}
Next, we introduce an intrinsic evaluation benchmark for
assessing distributional similarity between activities, independent of
downstream tasks.
First, we introduce the  experimental setup (\autoref{sec:intrinsic_idea}).
Then, we present the results of our evaluation
(\autoref{subsec:intrinsic_results}), before discussing them
(\autoref{sec:intrinsic_discussion}).

\subsection{Intrinsic Evaluation Setup}
\label{sec:intrinsic_idea}

Our intrinsic evaluation tests whether similar activities are also represented as similar by the evaluated method.
Since ground truth for activity similarity is not available, we simulate it by generating modified event logs in which we explicitly control which activities are similar. We refer to these as \emph{ground truth logs}.

\begin{figure}[b]
    \centering
\includegraphics[width=1\linewidth]{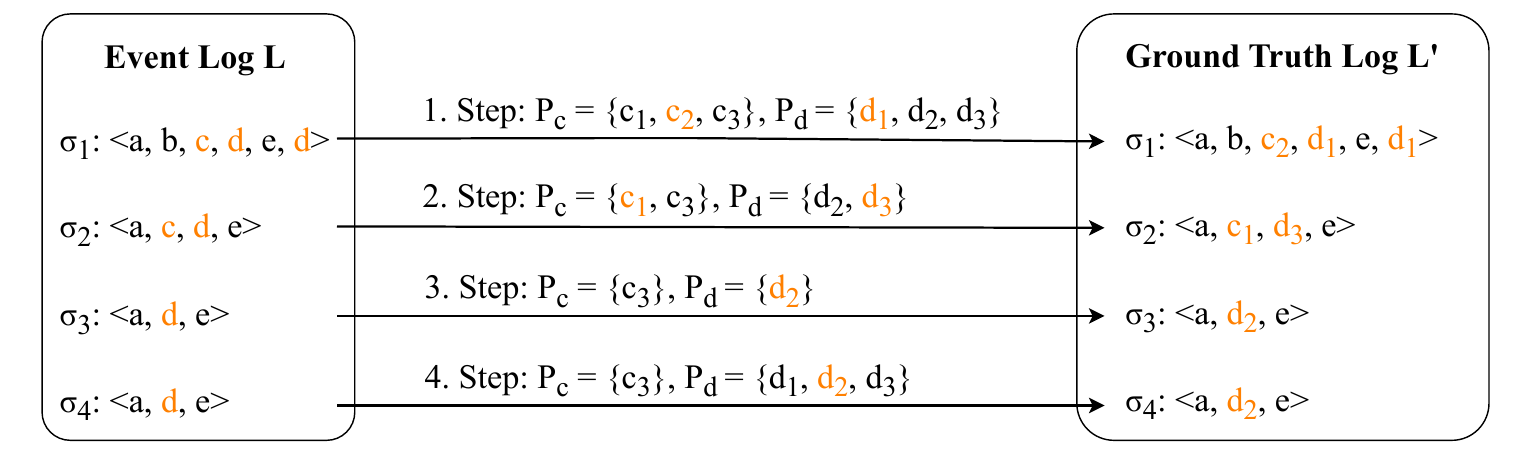}
    \caption{The procedure of creating a ground truth log $L'$ from $L=
    [\langle a,b,c,d,e,d\rangle, \langle a,c,d,e\rangle, \langle
    a,d,e\rangle^2]$ with $r=2$, $w=3$, and pools $P_c = \{c_1, c_2, c_3\}$ and $P_d =
    \{d_1, d_2, d_3\}$.}
\label{fig:Acitivity_Replacement}
\end{figure}

\mypar{Creation of Ground Truth Logs} \label{subsubsubsection:creation}
For each of the 28 real-world logs, we performed the following procedure. Given an event log \( L \), we randomly select \( r \) activities from the set of activities \( A_L \). Each selected activity is replaced by events drawn from a distinct pool of \( w \) new, unique activities not in \( A_L \), with \( w > 1 \). The \( w \) replacements for each original activity form a \emph{class}, while all other activities are considered \emph{out-of-class} to this \emph{class}. In each trace, all occurrences of a selected activity are replaced by the same randomly chosen replacement from its corresponding pool. Once a replacement is used, it is removed from the pool, which is reset when empty, ensuring balanced replacement across the log. 
Figure~\ref{fig:Acitivity_Replacement} illustrates an example where the two activities $c$ and $d$ ($r = 2$) are each replaced by three activities ($w = 3$).%

To assess the robustness of different distance measures under varying levels
of difficulty, we systematically generate ground truth logs of each original
event log. For each log, we consider all combinations of parameters $r \in
\{1, 2, ..., \min(|A_L|, 10)\}$ and $w \in \{2, 3, 4, 5\}$. These
parameters are chosen to ensure both computational feasibility and realistic
complexity. We start from $w = 2$ because at least two different activities
are needed to assess their proximity.

Given an event log $L$ and a combination of parameters $r$ and $w$, this yields $\binom{|A_L|}{r}$ possible ways to choose $r$ distinct activities. To ensure computational feasibility, we cap the number of generated logs per $(r, w)$ combination at a sampling size $s$ whenever $\binom{|A_L|}{r} > s$.
In this evaluation, we set $s$ to 5. This process leads up to $10 \text{
($r$ values)} \times 4 \text{ ($w$ values)} \times 5 \text{ (samples per
combination)} = 200$ new logs per original log.

\mycomment{
\mypar{Intrinsic Evaluation Measures} \label{subsubsec:intrinsic measures}
Given \( L' \), for each new activity \( a \in A(L') \setminus A(L) \), we
check:
\begin{enumerate}[label=(\Alph*)]
    \item Is \( a \) close to other activities that belong to the same class?
    \item Are there no activities \( a \in A(L') \) that are not from the
    same class closer to \( a \)?
\end{enumerate}
}

\mypar{Intrinsic Evaluation Measures} \label{subsubsec:intrinsic measures}
After constructing a ground truth log, we compute activity similarities for each evaluated method using the cosine distance between the returned activity embeddings, where a smaller distance indicates higher similarity. The only exception is the substitution scores approach~\cite{bose2009context}, where individual matrix cells are directly interpreted as similarity scores between activity pairs. Now, for each new activity \( a' \in A_{L'} \setminus A_{L} \), we want to determine:

\begin{enumerate}[label=(\Alph*)]
    \item whether \( a' \) is similar to the other new activities
     belonging to the same class, and
    \item whether activities that did not belong to the same class are not
    more similar to \( a' \) than those that did.
\end{enumerate}
\mycomment{
Given the set of activities, their embeddings, the quantified similarities,
and the true labels, we use the following four metrics proposed in the
literature. The first one evaluates aspect (A), while the
other three focus on aspect (B):
\begin{enumerate}
    \item \emph{normalized average diameter distance} $I_{comp}$~\cite{bose2009context}, to assess intra-class compactness;
    \item \emph{the nearest neighbor} $I_{nn}$~\cite{back2023comparing}, to check if the closest neighbor (in embedding space) belongs to the same class;
    \item \emph{precision@k} $I_{prec}$~\cite{back2023comparing}, to measure how many of the top-k retrieved activities are relevant (i.e., have the same label); and
    \item \emph{triplet measures} $I_{tri}$~\cite{back2023comparing}, to evaluate whether embeddings satisfy the triplet constraint: an anchor activity $a$ should be closer to an activity of the same class $\varphi(a)$ than to other activities $b \notin \varphi(a)$.
\end{enumerate}
}
Thus, we use the following four metrics. The first metric is used to evaluate aspect (A), while the remaining three evaluate aspect (B) and have been used in \cite{back2023comparing}. Let \(a', a'' \in A_{L'} \setminus A_{L} \), specifically, we evaluate:

\begin{enumerate}
    \item \emph{Compactness} ($I_{\text{comp}}$): For each class, we measure the average normalized pairwise similarity between all in-class activities.
    \item \emph{Nearest neighbor ($I_{\text{nn}}$)}: For each activity $a'$, we check whether its most similar activity belongs to the same class.
    \item \emph{Precision@k ($I_{\text{prec}}$)}: Given that $a'$ has $k$ other activities in its class, we measure how many of its $k$ most similar activities are from the same class.
    \item \emph{Triplet measure} ($I_{\text{tri}}$): 
    For each pair of in-class activities \(a', a'' \), we measure how often an out-of-class activity is less similar to \(a'\) than \(a''\) is to \(a'\).
\end{enumerate}

In our experiments, we average the evaluation results for each metric across all
generated logs and then average over all evaluated original logs.
Due to the excessively long runtime of the autoencoder-based method
\cite{gamallo2023learning}, there are no results where this approach did not
complete within 48 hours. %

\mycomment{
The experimental pipeline used for our intrinsic evaluation is depicted in
Fig.~\ref{fig:Experimental_Pipeline_Instrinsic}.

\begin{figure}[t]
    \includegraphics[width=1\linewidth]{IMG/05_Evaluation/Experimental_Pipeline_Instrinsic.pdf}
    \caption{Overview of the intrinsic evaluation pipeline. \todo{remove?}}
    \label{fig:Experimental_Pipeline_Instrinsic}
\end{figure}
}

\subsection{Intrinsic Evaluation Results}
\label{subsec:intrinsic_results}

Our results are shown in \autoref{tab:intrinsic_results}, where values in
the format x\,|\,y\,|\,z correspond to window sizes 3\,|\,5\,|\,9,
respectively; 
higher values are better for all four metrics. Note that each $I$ value reported is an average of approximately 200*28 logs; some logs have fewer than 10 unique activities, so we cannot generate all 200 combinations. 

\mypar{Our New Methods} Activity-activity co-occurrence matrices with sequential context, PMI
weighting, and a window size of 3 yield the best overall trade-off:
\[
   I_{\mathrm{comp}} = 0.83,\quad
   I_{\mathrm{nn}} = 0.75,\quad
   I_{\mathrm{prec}} = 0.72,\quad
   I_{\mathrm{tri}} = 0.83,
\]
outperforming all other methods on $I_{\mathrm{nn}}$ and $I_{\mathrm{tri}}$,
and ranking second in $I_{\mathrm{comp}}$ and $I_{\mathrm{prec}}$. The PPMI
variant under the same configuration matches $I_{\mathrm{nn}}$, but shows a
0.01 lower $I_{\mathrm{prec}}$ and a 0.02 lower $I_{\mathrm{comp}}$. Multiset-based methods achieve competitive
triplet scores but lag in $I_{\mathrm{nn}}$ and $I_{\mathrm{prec}}$.

Activity-context frequency matrices generally underperform compared to
activity-activity co-occurrence methods. The best configuration (sequential
context, PMI, window size 3) reaches $I_{\mathrm{nn}} = 0.74$,
$I_{\mathrm{prec}} = 0.73$, and $I_{\mathrm{tri}} = 0.78$, but suffers from
a substantially less compactness ($I_{\mathrm{comp}} = 0.55$).

\mypar{Existing Methods} All existing methods are outperformed by our new approaches on every metric except for $I_{\mathrm{comp}}$. The
embedding process structure approach \cite{chiorrini2022embedding}
achieves the best compactness ($I_{\mathrm{comp}} = 0.93$), but performs
poorly on $I_{\mathrm{nn}} = 0.52$ and $I_{\mathrm{prec}} = 0.48$. The
best-performing existing method, substitution scores \cite{bose2009context}
with window size 3, reaches $I_{\mathrm{nn}} = 0.61$, $I_{\mathrm{prec}} =
0.64$, and $I_{\mathrm{tri}} = 0.79$, but with a relatively low compactness
($I_{\mathrm{comp}} = 0.71$). The act2vec method \cite{de2018act2vec} performs slightly better than the substitution scores method on $I_{\mathrm{comp}}$ but worse on the other metrics. The autoencoder approach~\cite{gamallo2023learning} is the worst-performing method across all metrics. Results for the other methods, when only the logs used for the autoencoder are aggregated, are available in our GitHub repository. These results are consistent with those presented in the table. The previously mentioned activity-activity co-occurrence matrix with the best trade-off achieves on these logs: $I_{\mathrm{comp}} = 0.79$, $I_{\mathrm{nn}} = 0.70$, $I_{\mathrm{prec}} = 0.66$, and $I_{\mathrm{tri}} = 0.79$.

\mypar{Impact of Parameters} Sequential context consistently outperforms multiset context in our proposed
methods. However, for act2vec, the multiset-based bag-of-words architecture
performs slightly better than its sequential skip-gram variant. PMI postprocessing
improves $I_{\mathrm{nn}}$ and $I_{\mathrm{prec}}$, while slightly reducing
$I_{\mathrm{comp}}$ performance. Compared to PMI, PPMI typically yields equal
or lower scores but still outperforms the unweighted baselines. Across all
configurations, smaller window sizes yield better results, with window size
3 performing best.

\begin{table*}[t]
    \centering
    \setlength{\tabcolsep}{4pt}
    \renewcommand{\arraystretch}{1.2}
    \newcolumntype{P}[1]{>{\centering\arraybackslash}p{#1}}

    \caption{Intrinsic evaluation results, color-coded: \textcolor{red}{best}, \textcolor{blue}{2nd best}, \textcolor{green}{3rd best}; best existing method highlighted with \textcolor{violet}{violet}. Higher $I_{comp}$ $I_{nn}$, $I_{prec}$, and $I_{tri}$ are better. Values in the format x\,|\,y\,|\,z correspond to window sizes 3\,|\,5\,|\,9, respectively.}
    \label{tab:intrinsic_results}

\begin{threeparttable}
\begin{tabular}{|c|c|c|c|P{2cm}|P{2cm}|P{2cm}|P{2cm}|}
        \hline
        \parbox[t]{3mm}{\multirow{13}{*}{\rotatebox[origin=c]{90}{\textbf{Our New Methods}}}}
        & \textbf{Method} & \textbf{Context} & \textbf{PMI} & $I_{comp}$ & $I_{nn}$ & $I_{prec}$ & $I_{tri}$ \\
        \cline{2-8}
        & \multirow{6}{*}{\makecell{Activity-Activity\\Co-Occurrence\\Matrix}}
        & \multirow{3}{*}{Multiset} & No
        & 0.73\,|\,0.66\,|\,0.59 & 0.65\,|\,0.61\,|\,0.58 & 0.63\,|\,0.60\,|\,0.58 & \textcolor{green}{0.81}\,|\,0.73\,|\,0.68 \\
        \cline{4-8}
        & & & PMI
        & 0.8\,|\,0.76\,|\,0.75 & 0.70\,|\,0.66\,|\,0.63 & 0.68\,|\,0.66\,|\,0.62 & \textcolor{blue}{0.82}\,|\,0.75\,|\,0.70 \\
        \cline{4-8}
        & & & PPMI
        & 0.77\,|\,0.71\,|\,0.66 & 0.69\,|\,0.63\,|\,0.60 & 0.67\,|\,0.63\,|\,0.59 & \textcolor{blue}{0.82}\,|\,0.74\,|\,0.68 \\
        \cline{3-8}
        & & \multirow{3}{*}{Sequence} & No
        & 0.75\,|\,0.64\,|\,0.55 & 0.70\,|\,0.64\,|\,0.60 & 0.67\,|\,0.64\,|\,0.60 & \textcolor{blue}{0.82}\,|\,0.72\,|\,0.68 \\
        \cline{4-8}
        & & & PMI
        & \textcolor{blue}{0.83}\,|\,0.77\,|\,0.74 & \textcolor{red}{0.75}\,|\,0.70\,|\,0.65 & \textcolor{blue}{0.72}\,|\,0.69\,|\,0.64 & \textcolor{red}{0.83}\,|\,0.74\,|\,0.68 \\
        \cline{4-8}
        & & & PPMI
        & \textcolor{green}{0.81}\,|\,0.71\,|\,0.65 & \textcolor{red}{0.75}\,|\,0.68\,|\,0.63 & \textcolor{green}{0.71}\,|\,0.66\,|\,0.62 & \textcolor{red}{0.83}\,|\,0.73\,|\,0.66 \\
        \cline{2-8}
        & \multirow{6}{*}{\makecell{Activity-Context\\Frequency\\Matrix}}
        & \multirow{3}{*}{Multiset} & No
        & 0.6\,|\,0.51\,|\,0.45 & 0.63\,|\,0.58\,|\,0.57 & 0.63\,|\,0.58\,|\,0.56 & 0.69\,|\,0.62\,|\,0.59 \\
        \cline{4-8}
        & & & PMI
        & 0.56\,|\,0.39\,|\,0.27 & 0.69\,|\,0.62\,|\,0.58 & 0.68\,|\,0.61\,|\,0.57 & 0.73\,|\,0.65\,|\,0.60 \\
        \cline{4-8}
        & & & PPMI
        & 0.46\,|\,0.33\,|\,0.24 & 0.65\,|\,0.59\,|\,0.56 & 0.65\,|\,0.58\,|\,0.55 & 0.70\,|\,0.62\,|\,0.58 \\
        \cline{3-8}
        & & \multirow{3}{*}{Sequence} & No
        & 0.64\,|\,0.54\,|\,0.46 & 0.69\,|\,0.67\,|\,0.65 & 0.69\,|\,0.66\,|\,0.64 & 0.77\,|\,0.70\,|\,0.66 \\
        \cline{4-8}
        & & & PMI
        & 0.55\,|\,0.36\,|\,0.24 & \textcolor{blue}{0.74}\,|\,0.67\,|\,0.62 & \textcolor{red}{0.73}\,|\,0.66\,|\,0.62 & 0.78\,|\,0.70\,|\,0.64 \\
        \cline{4-8}
        & & & PPMI
        & 0.47\,|\,0.32\,|\,0.22 & \textcolor{green}{0.71}\,|\,0.64\,|\,0.61 & 0.70\,|\,0.63\,|\,0.60 & 0.76\,|\,0.68\,|\,0.63 \\
        \hline
        \multirow{3}{*}{%
          \raisebox{-9ex}{%
            \rotatebox[origin=c]{90}{%
              \shortstack{\textbf{Existing}\\\textbf{Methods}}%
            }%
          }%
        }
        & Substitution Scores \cite{bose2009context} & Sequence & - & 0.71\,|\,0.69\,|\,0.69 & \textcolor{violet}{0.61}\,|\,0.60\,|\,0.59 & \textcolor{violet}{0.64}\,|\,0.62\,|\,0.60 & \textcolor{violet}{0.79}\,|\,0.73\,|\,0.69 \\
        \cline{2-8}
        & \multirow{2}{*}{act2vec  \cite{de2018act2vec}} & Multiset & - & 0.75\,|\,0.74\,|\,0.74 & 0.58\,|\,0.57\,|\,0.55 & 0.57\,|\,0.55\,|\,0.54 & 0.76\,|\,0.76\,|\,0.75 \\
        \cline{3-8}
        & & Sequence & - & 0.72\,|\,0.71\,|\,0.70 & 0.56\,|\,0.54\,|\,0.52 & 0.55\,|\,0.53\,|\,0.51 & 0.78\,|\,0.77\,|\,0.76 \\
        \cline{2-8}
        & \makecell{\rule{0pt}{1em}Embedding Process\\Structure \cite{chiorrini2022embedding}\rule{0pt}{1em}} & Process Model & - & \textcolor{red}{0.93} & 0.52 & 0.48 & 0.77 \\
        \cline{2-8}
        & \makecell[r]{Autoencoder\cite{gamallo2023learning}\textsuperscript{*}} & Sequence & - & 0.38 & 0.07 & 0.06 & 0.45 \\
        \hline
    \end{tabular}
    \begin{tablenotes}
  \footnotesize
  \item[*] Computed over all logs except BPIC12, BPIC12 A, BPIC12 C, BPIC W C, BPIC17, BPIC18, BPIC19, Hospital Billing and RTFM (omitted due to >48 h runtime).
\end{tablenotes}
\end{threeparttable}
\vspace{-1em}
\end{table*}

\subsection{Intrinsic Evaluation Discussion}
\label{sec:intrinsic_discussion}

The results show a trade-off between context
specificity and generalization. Sequence‐based embeddings generally outperform multiset variants, underscoring the
importance of preserving activity order. Applying PMI helps to model what activities are similar and what are not.  PPMI tends to
over-smooth, and thus overly generalizes important information to characterize activity similarity.
Small windows (size = 3) consistently
outperform larger ones, indicating that immediate neighbors carry the most
important information, while distant activities introduce noise. The results of the embedding process structure approach show that its
representation of activities lacks expressiveness: while it places similar
activities close together (high $I_{\mathrm{comp}}$), it fails to distinguish
similar ones from dissimilar ones (low $I_{\mathrm{nn}}$ and
$I_{\mathrm{prec}}$). %

Overall, the results indicate that count-based approaches outperform neural network-based methods. Prior work on distributional similarity in NLP has shown that count-based models perform better on smaller datasets~\cite{sahlgren2016effects}. Since event logs are typically small by NLP standards, this may also explain their strong performance in our setting.

\section{Next Activity Prediction Evaluation}
\label{sec:next activity pred}

We now turn to the evaluation of the embeddings for an important
downstream task, i.e., next-activity prediction. Again, we first detail the
setup (\autoref{sec:downstream_idea}), before presenting the evaluation
results (\autoref{sec:downstream_results}), and their discussion
(\autoref{sec:downstream_discussion}).

\subsection{Next Activity Prediction Evaluation Setup}
\label{sec:downstream_idea}

Following the approach of \cite{gamallo2023learning}, we initialize the
embedding layer of a next-activity prediction model with pre-trained
activity embeddings and freeze its weights during training. 
After training, we then report classification accuracy for the next-activity prediction task as the ratio of correct to total predictions, i.e., $\mathrm{Accuracy} = \frac{\#\text{Correct Predictions}}{\#\text{Total Predictions}}$. In addition to overall accuracy, we evaluate performance on complex logs with a variant-to-trace ratio greater than $0.9$, i.e., $\frac{\#\text{Variants}}{\#\text{Traces}} > 0.9$. This subset includes the BPIC15 and NASA logs, which pose greater predictive challenges.

In this paper, we apply this method to the model proposed by Evermann et al. \cite{evermann2017predicting}, as it is reported to be the
most accurate next-activity prediction model that relies solely on activity
sequences as input, according to the survey by \cite{rama2021deep}. This
makes it an ideal benchmark for evaluating activity embeddings in isolation.
The model employs a single LSTM layer and predicts the next activity based
only on the sequence of preceding activities, without considering additional
event attributes such as timestamps or resources. While this simplicity
allows for fast training, it also results in comparatively lower prediction
performance, as noted in \cite{rama2021deep}. 

To provide a more robust comparison, we also include the model proposed by Tax et al. \cite{tax2017predictive}, which achieved the highest accuracy for next-activity prediction in the same survey. This model serves as a strong complementary baseline. Unlike Evermann's model, the Tax model performs joint prediction of the next activity and its timestamp, using activity one-hot encodings enriched with a five-dimensional time feature vector. 
The architecture consists of stacked LSTM layers but does not include an embedding layer for embedding evaluation.

For each event log, we apply a single holdout split of 64\% training, 16\%
validation, and 20\% test. Activity embeddings are pre-trained on the
combined training and validation sets and used to initialize the embedding
layer of the Evermann model, which is then frozen during training. We
benchmark next-activity prediction accuracy on the held-out test set using
the Evermann model with pre-trained (frozen) embeddings, the original
Evermann model with a randomly initialized (trainable) embedding layer, and
the Tax model. For this benchmark, we run experiments with 20 different
event logs. We excluded any logs where training the Evermann or Tax model
exceeded 48 hours.

\subsection{Next Activity Prediction Evaluation Results}
\label{sec:downstream_results}

The results of the next activity prediction evaluation are presented in \autoref{tab:nap_results}.

\mypar{Our New Methods}  
Activity-context frequency matrices consistently yield the best results, outperforming both activity-activity frequency matrices and all existing approaches. The highest performance is achieved with the PMI and PPMI configurations of the activity-context frequency matrices. Notably, our embeddings allow the simple Evermann model, without any temporal information, to approach the predictive performance of the state-of-the-art Tax model.

\mypar{Existing Methods}  
Overall, existing approaches perform at best on par with our methods. However, on more complex logs, even the strongest existing method, act2vec with multiset (CBOW), underperforms compared to our weakest new method, achieving only about half the performance of our best-performing approach.

\begin{table*}[t]
    \centering
    \setlength{\tabcolsep}{4pt}
    \renewcommand{\arraystretch}{1.2}
\newcolumntype{P}[1]{>{\centering\arraybackslash}p{#1}}
    \caption{Next activity prediction results, color-coded: \textcolor{red}{best}, \textcolor{blue}{2nd best}, \textcolor{green}{3rd best}; best existing method highlighted with \textcolor{violet}{violet}. Higher accuracy values are better. Runtime evaluation results in seconds, lower values are better. Values in the format x\,|\,y\,|\,z correspond to window sizes 3\,|\,5\,|\,9, respectively.}
    \label{tab:nap_results}

\begin{tabular}{|c|c|c|c|c|c|c|c|c|}
        \hline
        & \multirow{3}{*}{\textbf{Method}} & \multirow{3}{*}{\textbf{Context}} & \multirow{3}{*}{\textbf{PMI}} & \multicolumn{2}{c|}{\textbf{Next Activity Prediction}} & \multicolumn{2}{c|}{\textbf{Runtime Evaluation}} \\
        \cline{5-8}
        & & & & \textbf{Accuracy} & \makecell{\rule{0pt}{1em}\textbf{Accuracy}\\\textbf{Complex Logs}} & \textbf{BPIC15 1} & \textbf{BPIC19} \\

        \cline{1-8}
        \parbox[t]{3mm}{\multirow{12}{*}{\rotatebox[origin=c]{90}{\textbf{Our New Methods}}}} & \multirow{6}{*}{\makecell{Activity-Activity\\Co-Occurrence\\Matrix}}
        & \multirow{3}{*}{Multiset} & No
        & 0.51\,|\,0.52\,|\,0.54 & 0.23\,|\,0.24\,|\,0.22 & 3.23\,|\,3.47\,|\,3.89 & 1.41\,|\,1.60\,|\,2.12 \\
        \cline{4-8}
        & & & PMI
        & 0.56\,|\,0.56\,|\,0.57 & 0.26\,|\,0.25\,|\,0.25 & 5.79\,|\,5.90\,|\,6.31 & 1.52\,|\,1.61\,|\,2.12\\
        \cline{4-8}
        & & & PPMI
        & 0.56\,|\,0.56\,|\,0.55 & 0.26\,|\,0.26\,|\,0.24 & 5.87\,|\,6.07\,|\,6.44 & 1.39\,|\,1.73\,|\,2.27\\
        \cline{3-8}
        & & \multirow{3}{*}{Sequence} & No
        & 0.56\,|\,0.55\,|\,0.57 & 0.23\,|\,0.24\,|\,0.21 & 3.53\,|\,4.29\,|\,6.49 & 1.51\,|\,1.64\,|\,2.51 \\
        \cline{4-8}
        & & & PMI
        & 0.56\,|\,0.57\,|\,0.58 & 0.26\,|\,0.25\,|\,0.24 & 6.07\,|\,8.01\,|\,10.3 & 1.52\,|\,1.66\,|\,2.53 \\
        \cline{4-8}
        & & & PPMI
        & 0.56\,|\,0.57\,|\,0.57 & 0.26\,|\,0.25\,|\,0.24 & 6.01\,|\,7.02\,|\,9.73 & 1.39\,|\,1.66\,|\,2.55\\
        \cline{2-8}
        & \multirow{6}{*}{\makecell{Activity-Context\\Frequency\\Matrix}}
        & \multirow{3}{*}{Multiset} & No
        & 0.59\,|\,0.60\,|\,0.61 & 0.31\,|\,0.34\,|\,\textcolor{green}{0.35} & 3.65\,|\,21.59\,|\,15.25 & 1.49\,|\,1.53\,|\,2.23 \\
        \cline{4-8}
        & & & PMI
        & \textcolor{blue}{0.63}\,|\,\textcolor{green}{0.62}\,|\,\textcolor{red}{0.64} & \textcolor{green}{0.35}\,|\,\textcolor{blue}{0.36}\,|\,\textcolor{red}{0.38} & 7.75\,|\,32.74 \,|\,34.94 & 1.51\,|\,1.62\,|\,2.69 \\
        \cline{4-8}
        & & & PPMI
        & \textcolor{green}{0.62}\,|\,\textcolor{red}{0.64}\,|\,\textcolor{red}{0.64} & 0.34\,|\,\textcolor{red}{0.38}\,|\,\textcolor{red}{0.38} & 7.74\,|\,37.67 \,|\,35.27 & 1.50\,|\,1.63\,|\,2.80\\
        \cline{3-8}
        & & \multirow{3}{*}{Sequence} & No
        & 0.59\,|\
        ,0.61\,|\,0.61 & 0.30\,|\,0.33\,|\,0.34 & 3.69\,|\,14.80 \,|\,14.20 & 1.34\,|\,1.60\,|\,1.92\\
        \cline{4-8}
        & & & PMI
        & \textcolor{blue}{0.63}\,|\,\textcolor{red}{0.64}\,|\,\textcolor{red}{0.64} & \textcolor{blue}{0.36}\,|\,\textcolor{red}{0.38}\,|\,\textcolor{red}{0.38} & 9.02\,|\,37.34 \,|\,42.22 & 1.37\,|\,1.95\,|\,3.64 \\
        \cline{4-8}
        & & & PPMI
        & \textcolor{blue}{0.63}\,|\,\textcolor{red}{0.64}\,|\,\textcolor{red}{0.64} & \textcolor{blue}{0.36}\,|\,\textcolor{red}{0.38}\,|\,\textcolor{red}{0.38} & 8.96\,|\,34.83 \,|\,40.38 & 1.37\,|\,2.00\,|\,3.90 \\
        \hline
        \parbox[t]{3mm}{\multirow{7}{*}{\rotatebox[origin=c]{90}{\textbf{Existing Methods}}}} & Original Everman \cite{evermann2017predicting}& - & - & 0.50 & 0.09 & - & - \\
        \cline{2-8}
        & Substitution Scores \cite{bose2009context} & Sequence & - & 0.52\,|\,0.51\,|\,0.47 & 0.07\,|\,0.06\,|\,0.05 & 0.79\,|\,1.40 \,|\,2.80 & 1.40\,|\,1.61\,|\,3.15\\
        \cline{2-8}
        & \multirow{2}{*}{act2vec \cite{de2018act2vec}} & Multiset & - & \textcolor{violet}{0.57}\,|\,\textcolor{violet}{0.57}\,|\,0.55 & \textcolor{violet}{0.19}\,|\,0.18\,|\,0.18 & 7.46\,|\,7.73 \,|\,7.85 & 98.99\,|\,97.58 \,|\,101.19 \\
        \cline{3-8}
        & & Sequence & - & 0.54\,|\,0.53\,|\,0.51 & 0.15\,|\,0.14\,|\,0.13 & 12.06\,|\,14.23 \,|\,19.31 & 86.93\,|\,77.49 \,|\,76.57\\
        \cline{2-8}
        & Embedding Process Structure \cite{chiorrini2022embedding} & Process Model & - & 0.34 & 0.09 & 355.88 & 1.438\\
        \cline{2-8}
        & Autoencoder \cite{gamallo2023learning} & Sequence & - & 0.54 & 0.15 & 546.93 & 11207.91\\
         \cline{2-8}
         & Original Tax \cite{tax2017predictive} & - & - & 0.65 & 0.40 & - & -\\
        \hline
    \end{tabular}
\end{table*}

\subsection{Next Activity Prediction Evaluation Discussion}
\label{sec:downstream_discussion}

The strong performance of activity-context frequency embeddings in the Evermann model can be attributed to their richer representation. Unlike co-occurrence vectors, these embeddings explicitly encode contextual information, which becomes especially effective with larger window sizes (e.g., 9). Interestingly, for this task, multiset- and sequence-based contexts perform similarly. Although the autoencoder proposed by \cite{gamallo2023learning} is designed for complex processes, it performs worse than act2vec and significantly worse than our frequency-based methods. Our reimplementation closely matches the authors’ original results, indicating that the performance gap is not due to implementation or data differences. Rather, it highlights a limitation of the approach and the importance of diverse benchmarks: while the model performs reasonably well in next-activity prediction, its intrinsic similarity quality is poor. %

\section{Efficiency Evaluation}
\label{sec:runtime}

Finally, we explore the efficiency of the embedding methods. 
We first present the runtime required to derive the embeddings in \autoref{sec:runtime_idea} and \autoref{sec:runtime_results}, followed by the memory usage needed to store them \autoref{secc:memory}.

\subsection{Runtime Evaluation Experimental Setup}
\label{sec:runtime_idea}
We evaluated runtime performance by measuring the time required to generate activity embeddings and, for all approaches except the substitution scores method~\cite{bose2009context}, the time to compute pairwise cosine distances between embeddings. Experiments were conducted on two event logs selected for their statistics. The BPIC15 1 log with 398 activities, high variant ratio (0.98), long traces (avg. 43.55), and few traces (1,199). The BPIC19 has a huge trace count (251,734), but with a low trace variant ratio (0.05) and short traces (avg. 6.34).
For the process model embedding method \cite{chiorrini2022embedding}, we report runtimes without model discovery.

\subsection{Runtime Evaluation Results \& Discussion}
\label{sec:runtime_results}
The results of our runtime evaluation are also included in
\autoref{tab:nap_results}. We observe that count-based approaches are generally the fastest. But they take longer on logs with high behavioral variability (e.g., BPIC15 1), as matrix dimensionality increases. In contrast, prediction-based methods scale poorly with the number of traces, leading to longer runtimes on large logs like BPIC19. Activity-activity co-occurrence matrices and the substitution score approach are the fastest, completing in under 10 seconds on BPIC15 1 and under 3 seconds on BPIC19. However, substitution scores exclude the time needed to compute activity similarities. Activity-context frequency methods are slightly slower, especially with larger window sizes or PMI/PPMI, but still run within practical limits.

act2vec performs well on BPIC15 1 but is much slower on BPIC19. Process model embeddings are slow, particularly when including discovery (up to 585 seconds on BPIC15 1 and 34 seconds on BPIC19). The autoencoder is the most resource-intensive, taking over 3 hours on BPIC19.

\subsection{Memory Usage Evaluation}\label{secc:memory}

Activity-activity co-occurrence embeddings have dimensionality equal to the number of activities in the process. As processes typically involve a limited number of activities, the embeddings generally require very little memory by modern computational standards. For neural embeddings, memory usage is determined by the embedding dimension, which can be chosen by the user. Thus, when the number of activities is relatively small, such methods do not raise memory concerns and their space requirements are easily predictable once the number of activities is known in advance.

Memory usage is more critical for activity-context methods, where embedding dimensionality grows with the number of contexts. For sequential context interpretation with window size $w$, the upper bound is $|A_L|^w$~\cite{jurafsky2025speech}. Although this bound is unlikely to be reached in practice, due to the structured nature of most processes, high-dimensional vectors can still occur. The BPIC 2015 5 log produced 34,798-dimensional activity-context embeddings ($w=9$), totaling more than 110 megabyte for 398 activities at 64 bits per dimensional value.

\section{Conclusion}
\label{sec:Conclusion}

In this paper, we introduced count-based distributional similarity approaches that vary in context definition, matrix type, and post-processing. We also presented the first comprehensive benchmarking framework for activity similarity, covering intrinsic evaluation, downstream prediction performance, and computational efficiency. Across all benchmarks, our methods outperformed established baselines and showed strong efficiency in runtime and memory usage on typical event data.

Across the benchmarks, activity-activity co-occurrence and activity–context frequency-based methods perform similarly, except in next-activity prediction, where the latter excels at the cost of higher memory usage. Sequential context interpretation outperforms multiset; PMI consistently improves results; PPMI has only marginal 
effects; and the optimal window size is benchmark-dependent.

We note  that our intrinsic benchmark may be subject to potential biases introduced by the synthetic ground truth. Including real event logs with similar activities identified a priori by domain experts would strengthen the evaluation.
Furthermore, distributional similarity may fail in certain process
scenarios. Consider, for example, a medical
process for blood testing, where the sequence is largely fixed but includes
a branching point for the test result being either positive or negative. Distributional approaches would assign high similarity to these
activities, as they appear in nearly identical contexts. However, 
this similarity may be misleading: the 
activities following the branching point handle fundamentally different test outcomes.
Here, utilizing event attributes could address these issues.

Our work opens several directions for future research:
(i) further count-based approaches for
learning activity representations may be explored based on modern refinements, such as those
proposed in~\cite{levy2015improving}, including post-processing techniques, such as the shifted PMI;
(ii) an exploration how different hyperparameters, such as dimensionality of the embeddings can contribute to better performance; (iii) expanding our benchmark by considering a wider range of downstream tasks, such as anomaly detection or conformance checking.

\section*{Acknowledgment}
This research was partially funded by the German Federal Ministry of
Education and Research under grant 16DII133 (Weizenbaum Institute), the German Research Foundation (DFG) under grant 496119880, and an Erasmus+ grant.

\bibliographystyle{IEEEtran}
\bibliography{citations.bib}

\end{document}